\documentclass[useAMS,usenatbib]{mn2e}
\usepackage{aasmacros}
\usepackage{times}
\usepackage{graphicx}

\newcommand{\lcdm}[0]{${\Lambda}$CDM}

\newcommand{\fig}[1]{Fig.~\ref{#1}}
\newcommand{\eqn}[1]{equation~\ref{#1}}

\newcommand{\mnsec}[1]{Section~\ref{#1}}

\newcommand{\deltacf}[0]{$\xi(<\Delta)$}

\newcommand{\kms}[0]{\ensuremath{\mathrm{km\,s^{-1}}}}
\newcommand{\kpc}[0]{\ensuremath{\mathrm{kpc}}}
\newcommand{\wvunits}[0]{\ensuremath{\mathrm{kpc\,km^{-1}\,s}}}

\title[A halo star correlation function]{A
  two-point correlation function for Galactic halo stars}

\author[A.P. Cooper et al.]{A.P. Cooper$^{1,2}$\thanks{E-mail:
    acooper@mpa-garching.mpg.de}, S. Cole$^{1}$, C.S. Frenk$^{1}$, A. Helmi$^{3}$
  \\$^{1}$Institute for Computational Cosmology, Department of
  Physics, University of Durham, South Road, Durham, DH1 3LE, UK
  \\$^{2}$Max-Planck-Institut f\"{u}r Astrophysik,
  Karl-Schwarzschild-Str. 1, D-85748, Garching, Germany
  \\$^{3}$Kapteyn Astronomical Institute, University of Groningen, 
  P.O. Box 800, 9700 AV Groningen, Netherlands}
 
\begin{document}
  
\date{Accepted 2011 July 7. Received 2011 July 6; in original form 2010 November 9}

\pagerange{\pageref{firstpage}--\pageref{lastpage}} \pubyear{2011}

\maketitle

\label{firstpage}
\begin{abstract}

We describe a correlation function statistic that quantifies the
amount of spatial and kinematic substructure in the stellar halo. We
test this statistic using model stellar halo realizations constructed
from the Aquarius suite of six high-resolution cosmological N-body
simulations, in combination with the {\sc Galform} semi-analytic
galaxy formation model. The stellar haloes in the these simulations,
which form from disrupted satellites accreted between redshifts 1 and
7, show considerable scatter in the nature of their substructure. We
find that our statistic can distinguish between these different
realizations of plausible models for the global structure of the Milky
Way stellar halo. We apply our statistic to a catalogue of BHB stars
identified in the Sloan Digital Sky Survey. In our $\Lambda$CDM
simulations, we find examples of haloes with spatial and kinematic
substructure consistent with the available Milky Way data in the outer
halo. For the inner halo, the models predict stronger clustering than
observed, suggesting the existence of a smooth component, not
currently included in our simulations.

\end{abstract}


\section{Introduction}

In the Cold Dark Matter (CDM) cosmogony, galactic stellar haloes are
built up in large part from the debris of tidally disrupted satellites
(e.g. Searle \& Zinn 1978; White \& Springel 2000; Bullock \& Johnston 2005; Cooper {et~al.} 2010). Discovering
and quantifying halo structures around the Milky Way may provide a
useful diagnostic of the Galaxy's merger history
(e.g. Helmi \& de~Zeeuw 2000; Johnston {et~al.} 2008; G{\'{o}}mez \& Helmi 2010). Upcoming Milky Way
surveys (for example with PanSTARRS1, LAMOST, HERMES and the LSST)
will provide large datasets in which to search for structure. The
\textit{Gaia} mission will determine six-dimensional phase-space
coordinates for all stars brighter than $V\sim17$, from which it
should be possible to untangle even well-mixed streams in the nearby
halo (G{\'{o}}mez {et~al.} 2010).

Testing the CDM model by comparing these observations to numerical
simulations of stellar halo formation requires a straightforward
definition for the `abundance of substructure', one that can be
quantified with a method equally applicable to simulations and
observations. Algorithms already exist for identifying substructure in
huge multidimensional datasets (e.g. Sharma \& Johnston 2009), such as the
data expected from \textit{Gaia} supplemented by chemical abundance
measurements (Freeman \& Bland-Hawthorn 2002). These algorithms can
also be applied to simulations, although this is not
straightforward. One problem is that current (cosmological)
hydrodynamic simulations still fall short of the star-by-star
`resolution' of the \textit{Gaia} data, particularly in the Solar
neighbourhood (e.g. Brown, Vel{\'{a}}zquez \& Aguilar 2005).

In the outer halo, longer mixing times allow ancient structures to
remain coherent in configuration space for many gigayears. However, 6D
\textit{Gaia} data will be restricted to relatively bright stars. In
the near future, studies of the outer halo (beyond $\sim20$~kpc) will
continue to rely on a more modest number of `tracers' (giant and
horizontal branch stars). For these stars, typically only angular
positions and (more uncertain) estimates of distance and radial
velocity are available. In this regime, current simulations contain as
many particles as there are (rare) tracer stars in observational
samples, and can be compared directly with data that are already
available. Here we focus on quantifying the degree of structure in
rare tracer stars with a generic method, which we apply to
observational data and to simulations of Milky Way-like stellar
haloes.

Most studies of spatial and kinematic structure in the Milky Way halo
have given priority to the discovery of individual overdensities
(exceptions include Gould 2003, Bell {et~al.} 2008, Xue, Rix \& Zhao 2009,
Xue {et~al.} 2011 and Helmi {et~al.} 2011a). Relatively few
have investigated global statistical quantities for the entire stellar
halo, although several authors have suggested an approach based on
clustering statistics. Re~Fiorentin {et~al.} (2005) analysed the
velocity-space clustering of a small number of halo stars in the Solar
neighbourhood, using a correlation function statistic. Following early
work by Doinidis \& Beers (1989), Brown {et~al.} (2004) examined the angular
two-point correlation function of photometrically selected blue
horizontal branch (BHB) stars in the Two Micron All Sky Survey,
probing from $\sim2-9$~kpc. They detected no significant correlations
at latitudes $|b|\ga50\degr$, but did detect correlations on small
scales ($1\degr$, $\sim100$~pc) at lower latitudes, which they
attributed to structure in the thick disc. Lemon {et~al.} (2004) performed a
similar analysis for nearby F-type stars in the Millennium Galaxy
Catalogue and found no significant clustering.

Starkenburg {et~al.} (2009) used a correlation function in \textit{four}
dimensions to identify substructures in the Spaghetti pencil-beam
survey of the distant halo (Morrison {et~al.} 2000; Dohm-Palmer {et~al.} 2000).
With this method they obtained a significant detection of clustering
and set a lower limit on the number of halo stars in all
substructures. Similarly, Schlaufman {et~al.} (2009) constrained the mass
fraction of the halo in detectable substructure by estimating the
completeness of their overdensity detection
algorithm. Starkenburg {et~al.} and Schlaufman {et~al.}
concluded that $>10\%$ (by number of stars) and $\sim30\%$ (by volume)
of the Milky Way halo belongs to groups meeting their respective
definitions of phase space substructure. These methods were tested on
`mock catalogues' of tracer stars based on simplified models of the
stellar halo.

The work of Xue {et~al.} (2009, 2011) is particularly
  relevant. Xue {et~al.} (2009) considered the pairwise radial velocity
  separation of a sample of 2558 halo BHB stars as a function of their
  separation in space, but found no evidence of clustering. From
  comparisons to the simulations of Bullock \& Johnston (2005),
  Xue {et~al.} concluded that a pairwise velocity statistic was
  not capable of detecting structure against a more smoothly
  distributed background in phase space (made up from stars in
  phase-mixed streams). However, their observed signal was not
  compared to an expected signal from random realizations. More
    recently, Xue {et~al.} (2011) studied an enlarged catalogue of
    BHBs comprising more than 4000 stars. They quantified clustering
    in this larger sample using a four-dimensional metric
    (similar to that of Starkenburg {et~al.} 2009), finding a significant
    excess of clustering on small scales by comparison to smooth
    models. The conclusions of this more recent study by
    Xue {et~al.} agree with our own, as we discuss further
    in \mnsec{sec:conclusion}.

The statistic we develop in this paper also builds on the
  approach of Starkenburg {et~al.} (2009). We define a two-point correlation
  function based on a metric combining pairwise separations in four
  readily obtained phase space observables for halo stars (angular
  position, radial distance and radial velocity). We apply this
  statistic to the catalogue of BHB stars published by
  Xue {et~al.} (2008)\footnote{The larger sample used by Xue {et~al.} (2011)
    was not publicly available at the time this paper went to press.}
  and demonstrate that a significant clustering signal can be
  recovered from these data.

A clustering metric of the kind we propose can be tuned to probe a
specific scale in phase space by adjusting the weight given to each of
its components. In the stellar halo, however, many `components' may be
superimposed with a complex assortment of scales and morphologies in
phase space (Johnston {et~al.} 2008; Cooper {et~al.} 2010). For this reason, it is not
clear, a priori, what sort of signal to expect, or which scales are
most relevant. We find that we cannot identify an `optimal'
  metric. Instead, we make a fiducial choice based on the
self-consistent accreted halo models of Cooper {et~al.} (2010). These
incorporate an ab initio \lcdm{} galaxy formation model using
high-resolution cosmological N-body simulations from the Aquarius
project (Springel {et~al.} 2008).  We apply our fiducial metric to the
data and to these models. We find that even though both the metric and
our choice of scaling are simple, this approach has the power to
discriminate quantitatively between qualitatively different stellar
haloes.

We describe the basis of our method in \mnsec{sec:method} and the SDSS
DR6 BHB catalogue of Xue {et~al.} (2008) in \mnsec{sec:obsdata}. In
\mnsec{sec:sims} we describe our simulations
(\mnsec{sec:nbodygalform}) and our procedure for constructing mock
catalogues (\mnsec{sec:tracer_stars}). \mnsec{sec:fiducial} describes
our choice of a fiducial metric. In \mnsec{sec:segue} we apply our
method to quantify clustering in the SDSS data and compare this with
our mock catalogues. Our conclusions are given in
\mnsec{sec:conclusion}.

\section{A metric for phase-space distance}
\label{sec:method}

The most readily obtained phase-space observables for halo stars are
their Galactic angular coordinates, $l$ and $b$, heliocentric radial
distance, $r_{\mathrm{hel}}$, and heliocentric line-of-sight velocity,
$v_{\mathrm{hel}}$. Using its angular position and distance estimate,
each star can be assigned a three-dimensional position vector in
galactocentric Cartesian coordinates, $\bmath{r}\,(X,Y,Z)$, and a
radial velocity corrected for the Solar and Local Standard of Rest
motions, $v$. We begin by defining a scaling relation (metric),
$\Delta$, which combines these observables into a simple `phase-space
separation' between two stars:

\begin{equation}
{\Delta}_{ij}^2=
{|\bmath{r_i}-\bmath{r_j}|}^2+w_{v}^2({v_{i}-v_{j}})^2.
\label{eqn:delta_metric}
\end{equation}

Here, $|\bmath{r_i}-\bmath{r_j}|$ is the separation of a pair of stars
in coordinate space (in kiloparsecs), and $v_{i}-v_{j}$ is the
difference in their radial velocities (in kilometres per second). The
scaling factor $w_{v}$ has units of $\mathrm{kpc\,km^{{-}1}\,s}$, such
that $\Delta$ has units of kpc. The choice of $w_{v}$ is arbitrary
unless a particular `phase space scale' of interest can be
identified. This is not straightforward; we discuss some possible
choices below.
 
The aim of this paper is to explore $\xi({\Delta})$, the cumulative
two-point correlation function of halo stars in the metric defined by
Equation \ref{eqn:delta_metric}. Throughout, we use the
estimator \begin{equation} 1+\xi({\Delta}) =
  \frac{DD(<{\Delta})}{\left \langle RR(<{\Delta}) \right
    \rangle}.  \label{eqn:estimator}
\end{equation} Here $DD(<\Delta)$ counts the number of pairs in the
sample separated by less than $\Delta$, and $\left \langle RR(<\Delta)
\right \rangle$ is the equivalent count for pairs of random points
within the survey volume, averaged over several realizations. To
  generate these realizations we `shuffle' the data by randomly
  reassigning $(r_{\mathrm{hel}},v_{\mathrm{hel}})$ pairs to different
  $(l,b)$ coordinates drawn from the original catalogue\footnote{The
    same randomisation procedure was used by
    Starkenburg {et~al.} (2009). Xue {et~al.} (2011) use a similar
    procedure, but for each galaxy they re-assign $r_{\mathrm{hel}}$
    and $v_{\mathrm{hel}}$ \textit{separately} to different $(l,b)$
    coordinates.}.

Similar methods for quantifying the clustering of stars in a
  four-dimensional space are described by Starkenburg {et~al.} (2009, applied to a sample
    of giant stars from the Spaghetti survey) and
  Xue {et~al.} (2011, applied to a sample of BHB stars from
    SDSS). Our $\Delta$ metric is very similar to the
  $\delta_{4\mathrm{d}}$ metric of Starkenburg et al. in the limit of
  small angular separations\footnote{Starkenburg et al. developed
    their metric with the aim of identifying `true' pairs of stars
    with high confidence. In their definition (equation 1 of
    Starkenburg {et~al.} 2009), the $\delta_{4\mathrm{d}}$ separation
    between two stars depends not only on their actual phase-space
    coordinates, but also on how accurately those coordinates are
    measured. For example, moving two stars 10 kpc further apart and
    simultaneously decreasing the error on their distance measurements
    by a factor of 10 (relative to the average of the sample) results
    in the same $\delta_{4\mathrm{d}}$. Thus $DD/RR$ for separations
    in $\delta_{4\mathrm{d}}$ is not a straightforward measurement of
    physical clustering.}. We have verified that our metric gives
  similar results when we repeat the analysis of Starkenburg et
  al. using the Spaghetti sample of 101 halo RGB stars. For the rest
  of this paper we will focus on clustering in the SDSS DR6 BHB
  catalogue of Xue {et~al.} (2008).

\section{Observational data}
\label{sec:obsdata}

Xue {et~al.} (2008) have published a catalogue of 2558 stars from SDSS DR6,
which they identify as high-confidence halo BHBs (contamination
$<10\%$) using a combination of colour cuts and Balmer line
diagnostics. This sample ranges in distance from $4-60\,\mathrm{kpc}$;
a cut on distance error excluded more distant stars observed in
SDSS. Xue {et~al.} (2008) claim distance errors of $\sim10\%$ and radial
velocity errors of $5-20\,\mathrm{km\,s^{-1}}$. This catalogue is not
a complete sample of halo BHBs. In particular, Xue {et~al.} note
that the targeting of SDSS spectroscopy is biased against the
follow-up of more distant stars.

To isolate stars representing the kinematics of the halo (in order to
study the Galactic circular velocity profile) Xue {et~al.} (2008)
further restricted their sample to stars at heights
$|z|>4\,\mathrm{kpc}$ above the plane (avoiding the thick disc). We
also apply this cut (which excludes thick disc stars and low-latitude
halo stars alike), leaving 2401 stars in the sample. Finally we
exclude 9 stars in the sample identified by Xue {et~al.} (2009) as probable
globular cluster members. Thus, the sample against which we compare our
models contains 2392 stars from the 2558 stars in the Xue {et~al.} (2008)
catalogue. The effects of these refinements to the sample on the
recovered \deltacf{} signal are discussed in
\mnsec{sec:obsdata_clustering}.

\section{Stellar Halo Simulations}
\label{sec:sims}

\subsection{N-body and galaxy formation model}
\label{sec:nbodygalform}

The mock observations that we use to test the \deltacf{} correlation
function are derived from simulations of the accreted stellar halo
presented in Cooper {et~al.} (2010). These simulations approximate the
dynamics of stars in dwarf satellites of Milky Way-like galaxies by
`tagging' appropriate particles (i.e. those strongly bound within
subhaloes) in the Aquarius suite of high-resolution N-body simulations
(Springel {et~al.} 2008). Each `tag' associates a dark matter (DM)
particle with a particular stellar population of a given mass, age
and metallicity. This `tagging' technique is reasonable in the regime
of high mass-to-light ratios, which is supported in this case by
observations of stellar kinematics in dwarf galaxies
(e.g. Walker {et~al.} 2009).

The tagging method has a single free parameter, the fraction of the
most-bound particles chosen in each DM halo for each assignment of
newly-formed stars (for further details see Cooper {et~al.} 2010). The
value of this parameter (1 per cent) was fixed by requiring the
population of \textit{surviving} satellites (at the present day) to
have a distribution of half-light radius as a function of luminosity
consistent with Milky Way and M31 observations\footnote{The luminosity
  function of surviving satellites in these models also agrees with MW
  and M31 data. This agreement is mostly due to the underlying galaxy
  formation model.}. The Cooper {et~al.} models differ from the
earlier models of Bullock \& Johnston (2005) in that they treat the full
cosmological evolution of all satellites self-consistently in a single
N-body simulation, and use a comprehensive semi-analytic model of
galaxy formation (Bower {et~al.} 2006) constrained by data on large scales
and compatible with the observed MW satellite luminosity
function. Both the Cooper {et~al.} and the
Bullock \& Johnston simulations produce highly structured stellar
haloes built from the debris of disrupted dwarf galaxies. As we
discuss further below, other halo components formed {\it in situ} may
be present in real galaxies
(e.g. Abadi, Navarro \& Steinmetz 2006; Zolotov {et~al.} 2009; Font {et~al.} 2011) but these are expected
to be more smoothly distributed than the accreted component
(e.g. Helmi {et~al.} 2011a).

 \begin{figure*}
\includegraphics[height=60mmi,clip=True,trim=1cm 0cm 1cm 0 ]{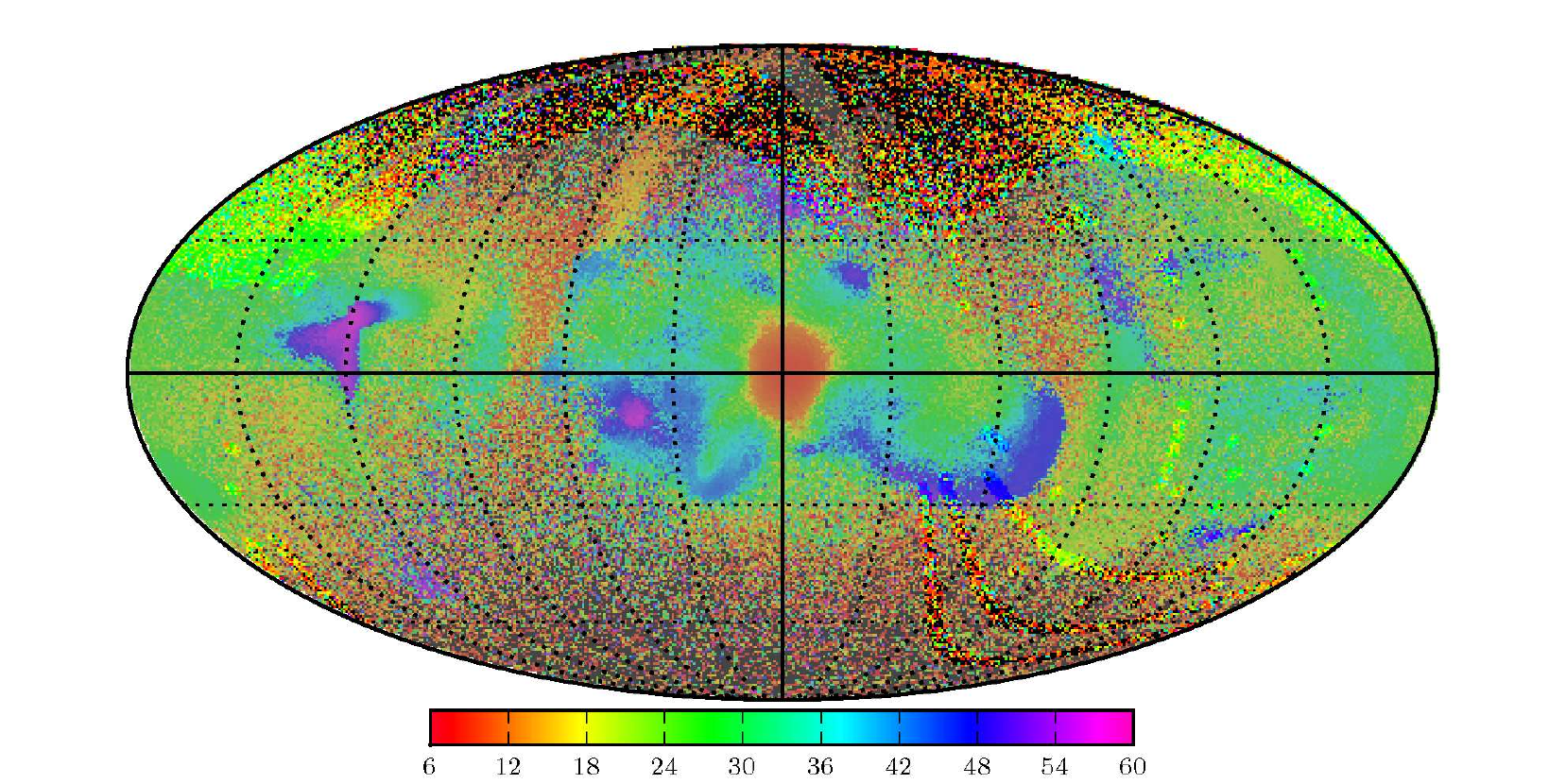}
\includegraphics[height=60mm,clip=True]{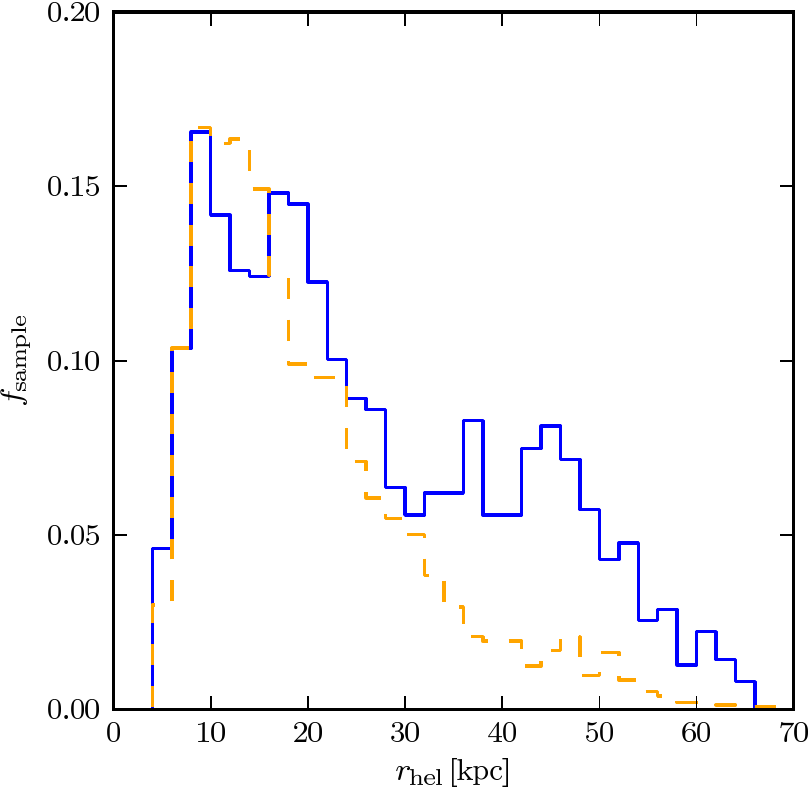}
\includegraphics[height=60mm,clip=True,trim=1cm 0cm 1cm 0 ]{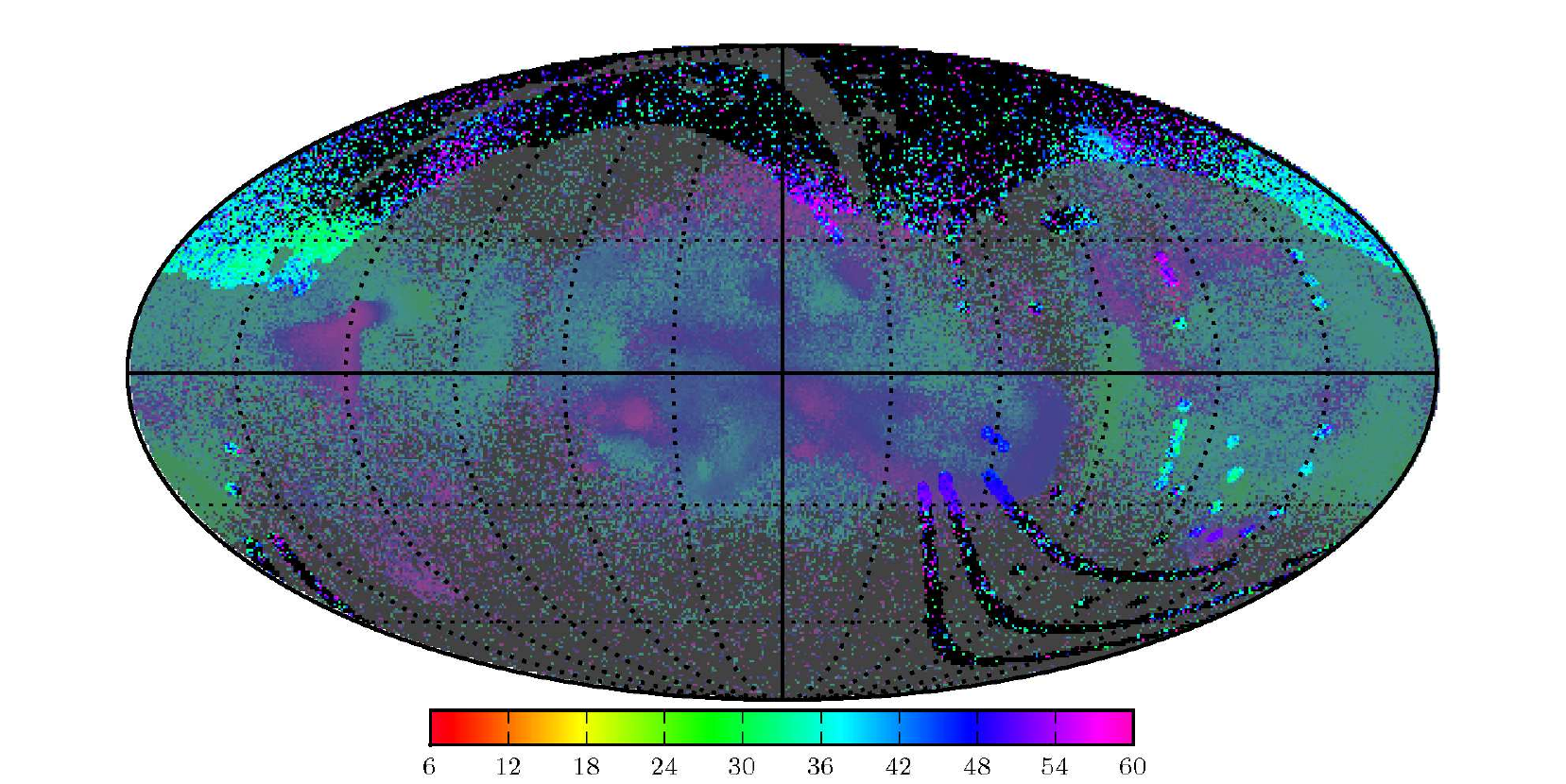}
\includegraphics[height=60mm,clip=True, trim=0 0 0 0 ]{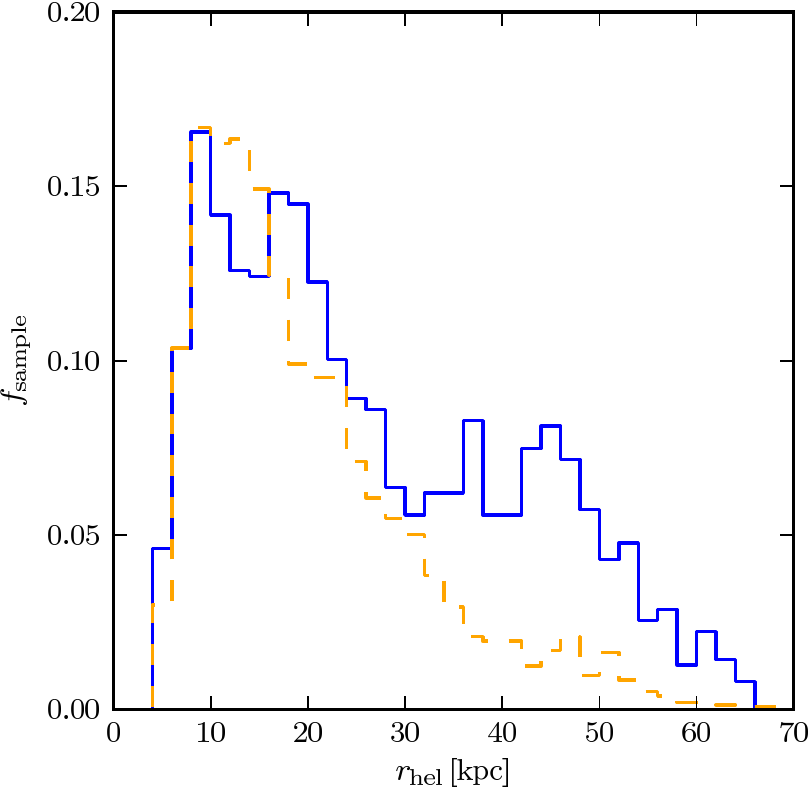}

\caption{\textit{Left panels:} An example of the sky distribution of
  halo BHB stars in Aq-A from the perspective of a `Solar' observer,
  shown as Mollweide projections in galactic coordinates centred on
  $(l,b)=(0,0)$. Colours indicate the mean heliocentric distance of
  stars in each pixel, in kiloparsecs. Pixels outside the SDSS DR6
  footprint are shown with lower contrast. The upper panel includes
  all BHB stars from $6$--$60$~kpc, the lower panel includes only
  stars in the range $30$--$60$~kpc. Our fiducial choice of the
  Galactic $Z$ axis with respect to the dark halo has been applied,
  but distances in these panels are not convolved with observational
  errors. The normalization of BHB stars per unit stellar mass in the
  halo has been increased in these panels to emphasise the
  distribution of structure. \textit{Right panels:} Blue histograms
  show the distribution of heliocentric distances (above) and
  heliocentric velocities (below) for the fiducial observer
  corresponding to the left panels, after convolution with
  observational errors (see text). Orange histograms are the
  distributions for stars in the Xue {et~al.} (2008) catalogue. To compare
  the shape of the two distributions, the normalization of the mock
  distributions in these panels has been matched to that of the
  observations for $r<20$~kpc, where the observations are most likely
  to be complete.}
\label{fig:fiducial}
 \end{figure*}

As in Springel {et~al.} (2008) and Cooper {et~al.} (2010), we refer to our
six simulations as haloes 
Aq-A, Aq-B, Aq-C, Aq-D, Aq-E and Aq-F. From these simulations, we
construct catalogues of tracer stars (BHB stars, for example) by
converting the stellar mass assigned to each dark matter particle into
an appropriate number of stars.

Each DM particle can give rise to many tracer stars if it is tagged
with sufficient stellar mass. We therefore interpolate the positions
and velocities of these tracer stars between nearby tagged DM
particles. To accomplish this, the 32 nearest phase space neighbours
of each tagged particle are identified using a procedure which we
describe below. The mean dispersion in each of the six phase-space
coordinates is then calculated for each particle as an average over
its neighbours. These dispersions define a 6D ellipsoidal Gaussian
kernel centred on the particle, from which the positions and
velocities of its tracer stars are drawn randomly. Each progenitor
galaxy (a set of tagged DM particles accreted by the main `Milky Way'
halo as members of a single subhalo) is treated individually in this
smoothing operation, i.e. particles are smoothed using only neighbours
from the same progenitor (so there is no `cross talk' between streams
from different progenitors). This procedure can be thought of as a
crude approximation to running our original simulation again including
each tracer star as a test particle.

%% L

The `distance in phase space' used to identify neighbours in the
interpolation scheme is defined by a scaling relation between
distances in configuration space and velocity space\footnote{In this
  part of the calculation, we are only interested in finding
  neighbours, so the absolute values of these distances are not
  important. This scaling of velocity space to configuration space for
  the purpose of resampling the simulations should not be confused
  with the $\Delta$ metric we define for our analysis of
  clustering.}. For each progenitor, we adopt an individual scaling
which corresponds to making the median pairwise interparticle
separation of its particles in configuration space (at $z=0$) equal to
their median separation in velocity space. In practice, the value of
this scaling parameter makes very little difference to the results we
present, when compared to the extreme choice of selecting only 32
velocity or position neighbours (disregarding the other three
coordinates in each case). Giving more weight to configuration-space
neighbours smears out velocity substructure within the debris of a
progenitor (for example, where two wraps of a stream pass near one
another). Giving more weight to velocity neighbours has the opposite
effect -- stars can be interpolated over arbitrarily large separations
in configuration space, but coherent velocity structures are
preserved. Therefore, the `optimal' choice is the scaling which
balances smoothing in configuration space against smoothing in
velocity space.

To quantify this balance between smoothing in configuration and
velocity space, we compute six smoothing lengths for each particle,
$\epsilon_{x,i}$ and $\epsilon_{v,i}$, where $i$ represents a single
dimension in space or velocity. To compute these dispersions,
  we use the 32 nearest phase-space neighbours of each particle. We
define the `optimum' choice of scaling for \textit{each} progenitor
galaxy as that which minimises the quantity

\begin{equation}
\sigma_{\epsilon}^{2} = \left (\frac{1}{\bar{\epsilon}_{x,\mathrm{min}}} \sum_{i=0}^{3}{\epsilon_{x,i}} \right )^{2} + \left ( \frac{1}{\bar{\epsilon}_{v,\mathrm{min}}} \sum_{i=0}^{3}{\epsilon_{v,i}}\right )^{2}.
\end{equation} This is the sum in quadrature of the mean smoothing lengths in
configuration and velocity space, normalized respectively by
$\bar{\epsilon}_{x,\mathrm{min}}$, the `minimal' mean smoothing length
in configuration space (obtained from the 32 nearest configuration space
neighbours) and $\bar{\epsilon}_{v,\mathrm{min}}$, the `minimal' mean
smoothing length in velocity space (obtained from the 32 nearest
velocity space neighbours). We find that the scaling obtained by
matching the median interparticle separations in position and velocity
as described above is typically a good approximation to this optimal
value -- a similar result is discussed in more detail by
Maciejewski {et~al.} (2009).

%% J

In the Cooper {et~al.} model, when a given stellar population is
formed, the most bound 1\% of DM particles in the corresponding dark
halo at that time are chosen as dynamical tracers of that
population. Hence, each DM \textit{particle} to which stars are
assigned has an individual mass-to-light ratio, M/L, which can be as
high as $\sim1$ (i.e. $M_{\mathrm{stellar}}\sim10^{4}\,M_{\sun}$) and
as low as $\sim10^{-6}$. This will affect the density of stars seeded
by a DM particle independently of the density of its neighbours in
phase space (i.e. a low M/L particle will create a denser `cluster' of
tracers relative to a high M/L particle with the same neighbouring
positions and velocities). We have tested an alternative approach in
which the M/L of each particle in a given progenitor is resampled by
distributing the total stellar mass of the progenitor evenly amongst
its tagged particles\footnote{This is almost equivalent to choosing
  M/L only once, at the time in the simulation when the progenitor
  falls into the main halo (similar to the lower-resolution model of
  De~Lucia \& Helmi 2008).}. We find that the extra clustering due to a
few `hot' particles in our default approach makes no difference to our
results.

%% A

\subsection{Tracer stars and mock observations}
\label{sec:tracer_stars}

Each N-body dark matter particle in our simulation contributes a
number of tracer stars to our mock observations, based on the stellar
population with which it has been `tagged'. In the specific case of
our comparisons to SDSS, these tracers are BHB stars meeting the selection
criteria of Xue {et~al.} (2008). Here we assume a global scaling between the
stellar mass associated with each N-body particle, $M_{\star}$, and
the number of BHBs it contributes to our mock catalogues, i.e.
$N_{\mathrm{BHB}} = f_{\mathrm{BHB}}M_{\star}$ where
$f_{\mathrm{BHB}}$ is the number of tracer stars per unit mass of the
original stellar population\footnote{We do this because we prefer to
  make a straightforward comparison with the observational data in
  this paper. In principle, the age and metallicity information
  associated with each stellar population in our model could be used
  to populate an individual colour-magnitude diagram for each N-body
  particle, and make a detailed prediction for the appropriate number
  of tracers. The `bias' of BHBs relative to the total stellar mass
  distribution of the halo (Bell {et~al.} 2010) may affect the
  clustering statistic recovered (Xue {et~al.} 2011), but this effect
  is beyond the scope of the present paper. }. For each N-body
particle, the actual number of BHB stars generated is drawn from a
Poisson distribution with mean $N_{\mathrm{BHB}}$. The correlation
function results described below are not sensitive to the choice of
$f_{\mathrm{BHB}}$, provided that the underlying distribution is
well-sampled at a given scale. We have therefore selected a fiducial
value of $f_{\mathrm{BHB}}^{-1} = 3000\,M_{\sun}/\mathrm{star}$. In
creating the mock catalogue, we do not include any stars
gravitationally bound to satellites. However, we do include stars in
their tidal tails (which, by our definition, are part of the stellar
halo).

Using our simulated BHB catalogues, we create mock observations for
comparison to the Xue {et~al.} (2008) data as follows. First we located the
observer at a radius $r_{\sun}=8\,\mathrm{kpc}$ from the centre of the
halo. For our main comparison to the data, we restrict all observers
to the same `Galactic plane', with each random vantage point differing
only in its azimuthal angle in this plane and in the `polarity' of the
Galactic rotation axis (the $Z$ coordinate). However, the orientation
of the rotation axis cannot be directly constrained by the simulation,
which only models the \textit{accreted} component of the halo and the
bulge, and not the in situ formation of a stellar disc. As described
in Cooper {et~al.} (2010), the accreted `bulge' is prolate or mildly
triaxial. We define the minor axis of this bulge component
(conservatively defined by all accreted stars within $r<3$~kpc;
Cooper {et~al.} 2010) as the Galactic $Z$ axis. This axis is essentially
identical to the minor axis of the dark halo within $r<3$~kpc. There
are other plausible choices of Galactic plane (for example, relative
to the shape or spin vectors of the entire dark halo, rather than the
stars in its inner regions). However, any choice is somewhat arbitrary
without a self-consistent simulation of disc formation\footnote{In a
  full hydrodynamic simulation the effects of feedback and adiabatic
  contraction may also make the dark halo itself more spherical
  (e.g. Tissera {et~al.} 2010; Abadi {et~al.} 2010).}.

Having chosen a location for the observer, we select all tracer stars
within the spectroscopic footprint of SDSS DR6 having galactocentric
distance in the range 20--60~kpc (our principal comparison will focus
on the outer halo as defined by this distance range, although we also
study the ranges 5--60~kpc and 5--20~kpc below). Galactic longitude
and latitude are defined such that $(l,b)=(0,0)$ is the vector
directed from the observer to the centre of the halo. We set the
heliocentric velocity components of each star assuming a Solar motion
of $U,V,W = (10.0,5.2,7.2)\:\mathrm{km\,s^{{-}1}}$ (Dehnen \& Binney 1998) and
a velocity of the Local Standard of Rest about the Galactic centre
$v_{\mathrm{LSR}}=220\mathrm{\:km\,s^{{-}1}}$. We compute $(X,Y,Z)$
and $v_{\mathrm{los}}$ as described by Xue {et~al.} (2008). Finally,
distances and velocities are convolved with Gaussian observational
errors of $\sigma_{d}= 10\%$ and $\sigma_{v}=15\;\mathrm{km\,s^{-1}}$
respectively (Xue {et~al.} 2008).

In both the mock observations and the real data, the average random
pair count $\langle RR \rangle$ is calculated by reshuffling distances
and velocities among the positions on the sky of stars many times. We find that
by using 500 random catalogues to calculate $\langle RR \rangle$ for
each mock observation and performing 500 mock observations in each
halo, we obtain a sufficiently well-converged estimate of the
distribution of \deltacf{}.

\fig{fig:fiducial} illustrates the structure of one of our haloes and
verifies that our mock observations can result in distributions of
heliocentric distance and heliocentric radial velocity similar to the
SDSS data of Xue {et~al.} (2008). In this figure we have specifically chosen
an observer orientation in halo Aq-A such that the distributions of
distance and velocity we recover are close to those of the data, after
convolution with typical observational errors. We have aligned
the Galactic $Z$ axis of the mock observer with the minor axis of 
the dark halo, as described above. This confines the most prominent
structures in the stellar halo to low Galactic latitudes, outside the
SDSS DR6 spectroscopic footprint. Of course, the simulated haloes are
inhomogeneous on large scales, and there are many choices of observer
in each halo that \textit{do not} resemble the SDSS data\footnote{As
  discussed by Xue {et~al.} 2008, the completeness of the data declines
  at larger distances ($r_{\sun}>20$~kpc). Mock catalogue distributions
  that match the observed distributions well at $r_{\sun}<20$~kpc
  typically show a flatter profile with identifiable overdensities
  (streams) at larger distances. As the SDSS spectroscopic selection
  function for the data we use is difficult to quantify (Xue {et~al.} 2008),
  we do not explore the effects of incompleteness in this paper.}.

\section{Distance - velocity scaling in the $\Delta$ metric}
\label{sec:fiducial}

\begin{figure}
\includegraphics[width=84mm,clip=True]{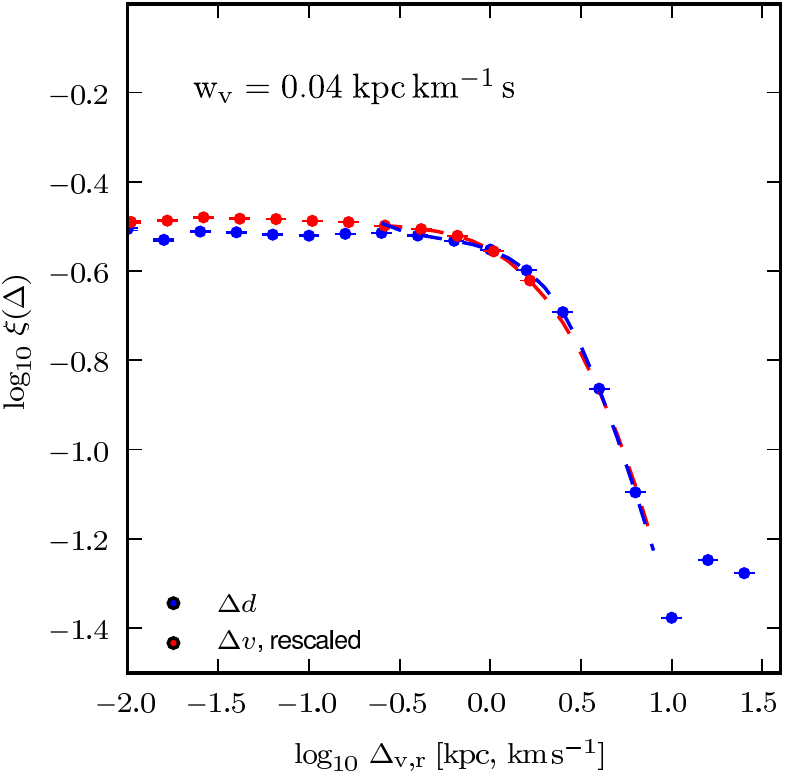}
\caption{Correlation functions in spatial separation (blue) and velocity
  separation (red) for stars in halo Aq-A. The velocity separation
  correlation function has been scaled by a factor
  $w_{v}=0.04\,\wvunits$ to match the turnover in the configuration
  space separation correlation function.}
\label{fig:accretedscale}          
\end{figure}

Before \deltacf{} can be computed, a value must be chosen for the
  velocity-to-distance scaling $w_{v}$ in Equation
  \ref{eqn:delta_metric}. There is no clearly well-motivated way to
  choose this value, and in the absence of a physical justification,
it must be treated as a free parameter. The choice of $w_{v}$
determines the scale of substructure to which the correlation function
is most sensitive. Naively, we expect this to be the typical width and
transverse velocity dispersion of a `stream'. It is preferable to fix
this parameter in a universal manner that does not depend on the
  details of a particular survey. We make a fiducial choice of
$w_{v}$ as follows.

In each simulated halo we adopt the SDSS-like survey configuration
described in \mnsec{sec:tracer_stars} (without observational errors or
assumptions about the orientation of the Galactic plane). We
construct one dimensional distributions of the separation in radial
distance and velocity between stars. We generate many random
realizations of these distributions by first convolving each simulated
star with Gaussian smoothing kernels of width $8\,\kpc$ (distance) and
$80\,\kms$ (velocity), and then drawing randomly from these `smoothed'
distributions. The smoothing scales were chosen as a compromise
between signal (diminished by oversmoothing) and noise (increased by
undersmoothing). Using these random realizations we construct
one-dimensional correlation functions for each distribution. These two
correlation functions are shown for halo Aq-A in
\fig{fig:accretedscale}. Although the signals are intrinsically weak,
they have a very similar shape for both distributions, each with a
characteristic `turnover' scale. Matching this scale in the two
correlation functions corresponds to $w_{v}\sim0.04\pm0.01\,\wvunits$
for the six haloes, which we adopt as a fiducial value. We caution
that although the scales on which we match the one-dimensional
correlation functions are somewhat smaller than the smoothing scales
we adopt to create the random distributions, this does not guarantee
that our choice of $w_{v}$ is unaffected by our choice of smoothing.

Clearly, there are other ways of fixing $w_{v}$. In practice, however,
our conclusions are not highly sensitive to the value of this
parameter. Values of the order of $w_{v}\sim0.01$--$1.0\,\wvunits$
result in very similar $\xi({\Delta})$ correlation functions. Values
lower than $0.01\,\wvunits$ recover very little signal. Values above
$1\,\wvunits$ treat $1\,\kms$ velocity differences as equivalent to
$>1\,\kpc$ separations in space, and so make the cumulative
correlation function very noisy on small scales for only a marginal
increase in the overall signal. (This noise, in turn, increases the
scatter between signals measured by different observers.) We find that
our choice of $w_{v}\sim0.04\,\wvunits$ is a reasonable
compromise. Our method for choosing $w_{v}$ can be compared with that
of Starkenburg {et~al.} (2009), who determine the equivalent of $w_{v}$ in
their metric to be the ratio of the Spaghetti survey limits in radial
distance and velocity ($0.26\,\wvunits$). Either value is acceptable
to illustrate our approach and compare to simulations. We therefore
adopt $w_{v}\sim0.04\,\wvunits$.

\section{Clustering of SDSS BHB stars}
\label{sec:segue}

\subsection{Clustering in the Xue et al. sample}
\label{sec:obsdata_clustering}

\begin{figure}
\includegraphics[width=84mm,trim= 0.0cm 0cm 0cm 0.0cm,
  clip=True]{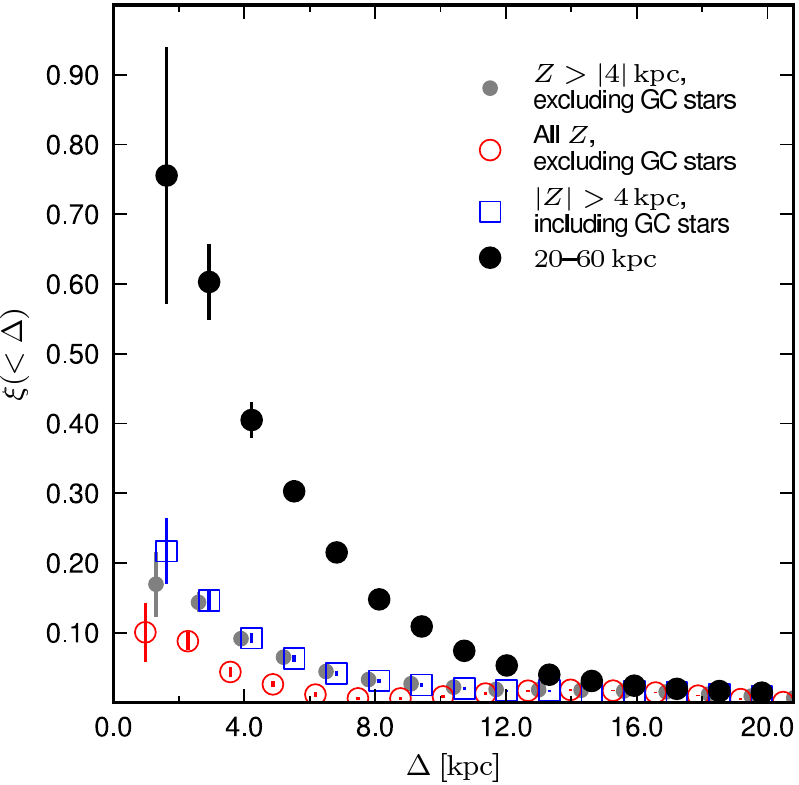}
\caption{\deltacf{} correlation function for the SDSS BHB sample of
  Xue {et~al.} (2008). Black points (with Poisson error bars) show
  \deltacf{} computed for all stars at galactocentric distances
  greater than 20~kpc. Grey points show the result for all stars in
  the main sample (galactocentric distances of 4--60~kpc). The blue
  squares include 9 stars suspected to belong to globular clusters,
  while red circles include stars at low Galactic $|z|$ heights
  (possible thick disc stars). Neither of these contributions are
  relevant for the more distant selection shown by the black points.}
\label{fig:sdss_real}          
\end{figure}

% Separate into 20kpc and <20kpc

\fig{fig:sdss_real} shows \deltacf{} computed for 2392 BHB stars in
the Xue {et~al.} (2008) sample (\mnsec{sec:obsdata}; grey points). Stars at
small separations in the metric of \eqn{eqn:delta_metric}
($\Delta<4\,\mathrm{kpc}$) show significant clustering. The amplitude
of the signal increases if we restrict the sample to larger
galactocentric distances, $r>20$~kpc (black points). At larger
distances substructure is expected to be dynamically young and to have
undergone less phase mixing. Our finding of stronger clustering for
more distant halo stars is in qualitative agreement with the results
of Xue {et~al.} (2011).

Although we appear to recover a significant clustering signal in
\fig{fig:sdss_real}, we have only one SDSS survey. The observed signal
may be an artifact of the particular structures covered by the SDSS
footprint. Other parts of the halo may be smoother or more structured,
or may appear so when viewed from different points around the Solar
circle. We will address this issue of sample variance in the following
section using our mock catalogues.

We show two further permutations of the Xue {et~al.} (2008) sample in
\fig{fig:sdss_real}. The first of these (red open circles) includes
stars close to the Galactic plane, $|Z|<4\,\mathrm{kpc}$. These were
excluded from the main sample of Xue {et~al.} (2008) to excise the thick
disc.

Although only $\sim150$ stars are excluded by the cut on
  $|Z|$, they make a substantial difference to the correlation
  function, suppressing the clustering signal on scales below
  $\Delta<\sim8\,\mathrm{kpc}$. In the SDSS data the majority of
  low-$|Z|$ stars are at small heliocentric radii. These stars
  constitute a foreground `screen' with a relatively smooth
  distribution, which may dilute the signal of correlated stars.

The final sample shown in \fig{fig:sdss_real} (blue open squares)
includes all stars from the main sample (grey points) and a further
nine BHB stars identified as globular cluster members by
Xue {et~al.} (2009). Two of these are from one cluster, and seven from
another. Including these stars marginally increases the clustering
signal in the smallest-separation bin. This shows that the technique
is sensitive to the clustering of stars on these scales, which
correspond to separations comparable to the distance and velocity
errors of the data.

\subsection{Comparison with Mock Catalogues}
\label{sec:mock_clustering}

\begin{figure}
\includegraphics[width=84mm, clip=True]{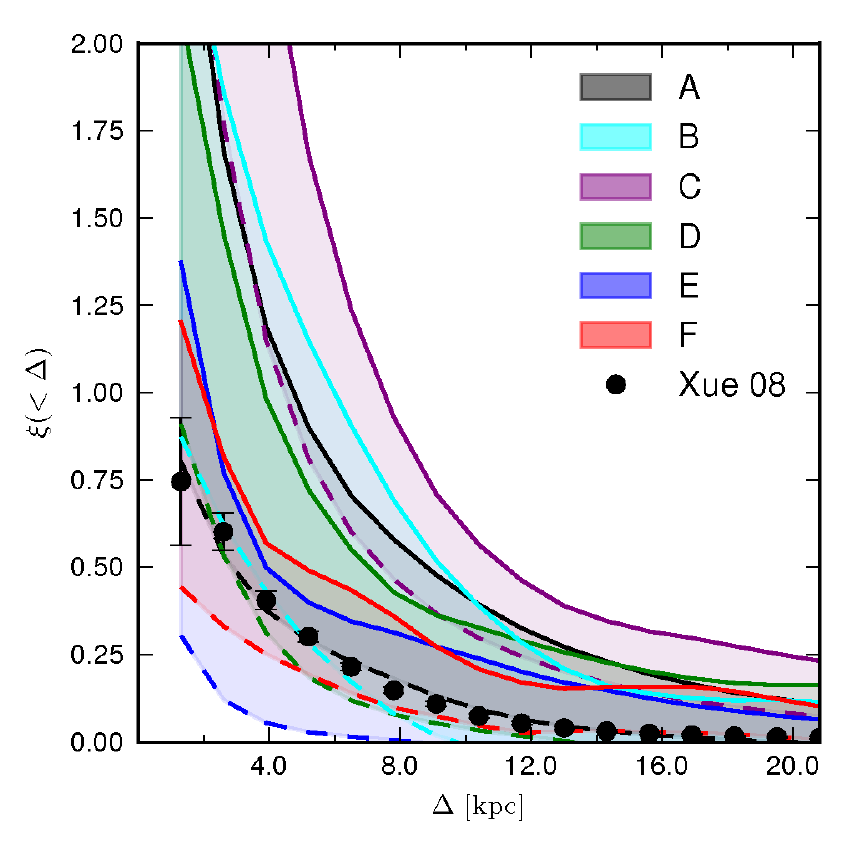}
\caption{\deltacf{} for halo stars of galactocentric distances
  $20<r<60$~kpc, computed for 500 observers in the six Aquarius
  simulations of Cooper {et~al.} (2010). Coloured regions (delineated by
dashed and solid lines) indicate the
  10--90\% range of \deltacf{} in each bin of $\Delta$, for each
  halo. All observers are restricted to the Galactic plane, and
  distances and velocities are convolved with observational errors (see
  text). Black points with error bars reproduce the observed BHB
  \deltacf{} shown in \fig{fig:sdss_real}.}
\label{fig:sdss_mock_pa}          
\end{figure}

In \fig{fig:sdss_mock_pa} we show the range of \deltacf{} measured for
each of the six simulations described in \mnsec{sec:sims},
overplotting the data shown in \fig{fig:sdss_real}. As described in
\mnsec{sec:tracer_stars} these results correspond to mock BHB
catalogues for 500 Solar observers located in the `Galactic plane',
with typical observation errors in distance and velocity. The dashed
and solid lines bounding each coloured region in
\fig{fig:sdss_mock_pa} correspond to the 10th and 90th percentile
values we obtain in each bin of \deltacf{}. At a given scale in our
$\Delta$ metric, the strength of the clustering signal varies
considerably from halo to halo and between individual observers.

The overall trend of \deltacf{} is similar to the observations in all
haloes, although the clustering signal rises more steeply on small
scales in most of the simulations. On scales $\Delta < 4$~kpc, significant
clustering is detected by all observers in five of the six haloes. The
exception is halo Aq-E (dark blue). This stellar halo is highly
concentrated, and at $r>20$~kpc is dominated by a single radial stream
(see figures 6 and 7 of Cooper {et~al.} 2010).

We find that two haloes, Aq-E and Aq-F (red), are consistent with the
observed \deltacf{} on all scales. The structure of Aq-F is atypical
for the sample -- most of its stars are accreted in a late 3:1 merger
and its surface brightness at the Solar radius is substantially higher
than current estimates for the Milky Way halo. In projection, Aq-F
resembles the `shell'-dominated haloes observed in a number of nearby
elliptical galaxies. Meanwhile haloes Aq-A (black), Aq-B (cyan) and
Aq-D (green) are marginally inconsistent with the data: below
$\Delta\sim4$~kpc, $\sim90$ per cent of mock observations in these
haloes imply a greater degree of clustering than we find for the Milky
Way, particularly on small scales. Aq-C (purple) is entirely
inconsistent with the Milky Way observations on all scales, showing a
much higher degree of clustering. Beyond $20$~kpc, the sky of an
observer in Aq-C is dominated by two bright tidal streams on wide
($\sim100$~kpc) orbits. Although their orbital planes are
approximately coincident with our definition of the Galactic plane,
nevertheless sections of these streams intrude on the SDSS footprint
at low Galactic latitudes.

\begin{figure*}
\includegraphics[height=65mm,clip=True]{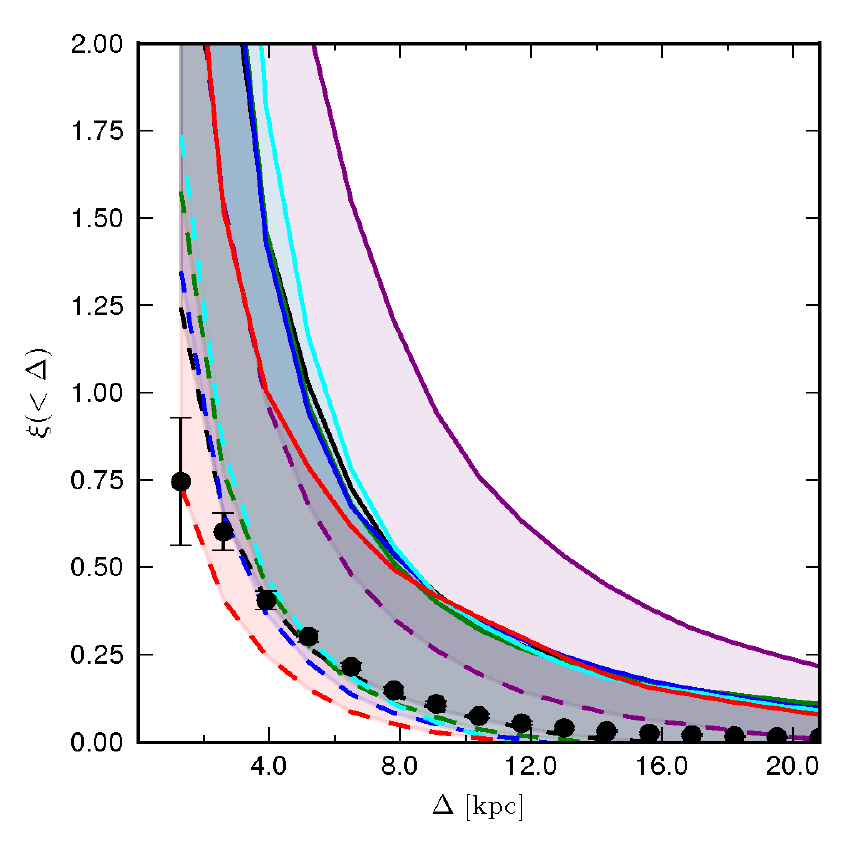}
\includegraphics[height=65mm,trim=1.3cm 0 0 0,clip=True]{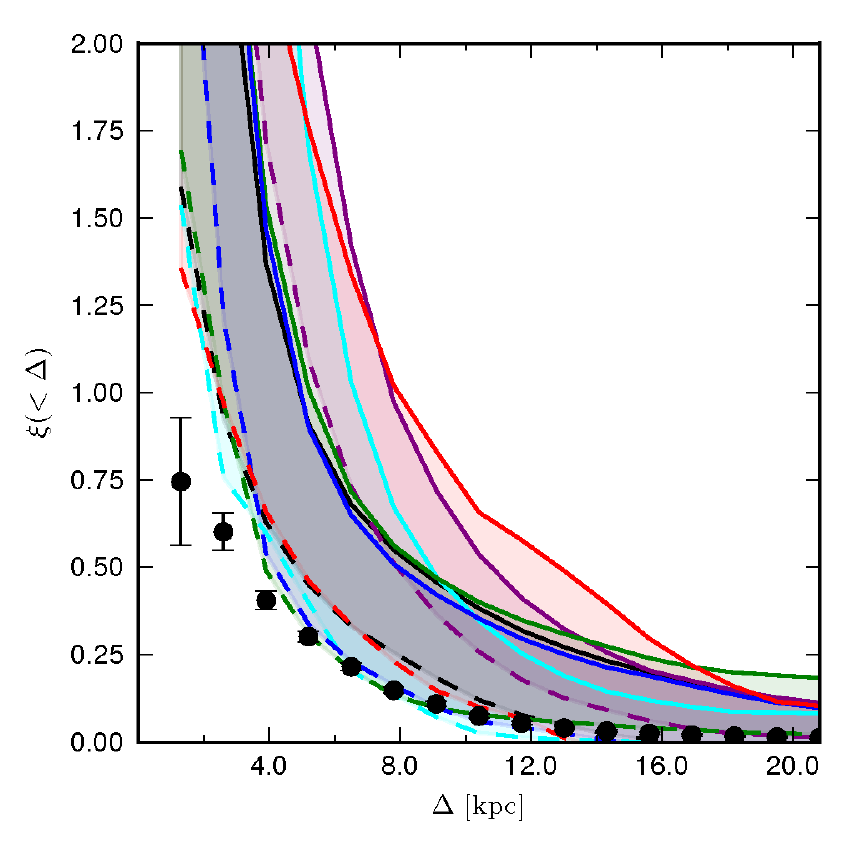}
\includegraphics[height=65mm,trim=1.3cm 0 0 0,clip=True]{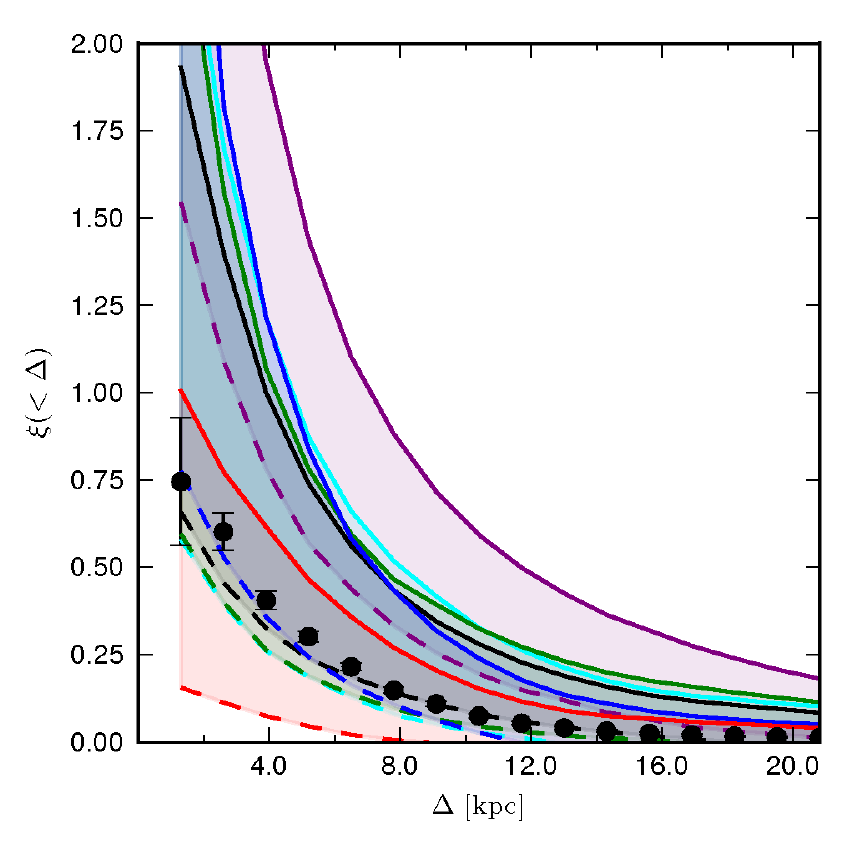}
\caption{\deltacf{} for mock observations, as
  \fig{fig:sdss_mock_pa}. From left to right, we vary our modelling
  assumptions as follows: (a) no restriction on the alignment of the
  Galactic plane with respect to the dark halo (the observer is
  located randomly on a sphere of radius $r_{\sun}=8$~kpc) and no
convolution of the data with the expected observational errors; (b) the
  observer is restricted to the Galactic plane as in
  \fig{fig:sdss_mock_pa}, but the mock observations are \textit{not}
  convolved with expected observational errors; (c) as panel (a), but
  mock observations \textit{are} convolved with errors. Colours are as
  \fig{fig:sdss_mock_pa}}
\label{fig:sdss_mock_no}          
\end{figure*}

The DR6 footprint and the cut on extra-planar height in the
Xue {et~al.} (2008) sample exclude stars near the Galactic plane from our
clustering analysis. \fig{fig:sdss_mock_no} illustrates how our
definition of the Galactic plane influences the halo clustering
signal. In panel (a) the orientation of the Galactic plane with
respect to the halo is chosen randomly for each of the 500 mock
observers (i.e. observers are distributed over a sphere of radius
$r_{\sun}=8$~kpc), whereas in panel (b) the galactic Z direction
is aligned with the minor axis of the halo for all observers as in
\fig{fig:sdss_mock_pa}. To focus on the effects of this alignment, the
distances and velocities of stars in these two panels have
\textit{not} been convolved with observational errors.

The systematically higher clustering signals in panel (b) of
\fig{fig:sdss_mock_no} suggest that the plane perpendicular to the
minor axis of the dark matter halo is special. In Cooper {et~al.} (2010) and
above, we have noted the strong correlation between the shape of the
dark halo and the inner regions of the stellar halo. This alignment of
halo structure also extends, more loosely, to other prominent stellar
halo structures at large distances. An overall flattening of the
stellar halo arises because our dark matter haloes are embedded in
long-lived filaments of the cosmic web. Typically one or two such
filaments dominate the infall directions of both satellite galaxies
and smoothly accreted dark matter, which also contributes to the shape
of the dark halo
(e.g. Libeskind {et~al.} 2005; Lovell {et~al.} 2011; Wang {et~al.} 2011; Vera-Ciro {et~al.} 2011). The
distribution of stars stripped from infalling satellites echoes the
large-scale correlation of their orbital planes.

Because of this flattened global structure, the distribution of halo
stars in our choice of Galactic plane tend to be more smoothly
distributed (i.e. this plane contains more diffuse phase-mixed
material as well as coherent substructure). Panel (b) demonstrates how
the `contrast' of small scale substructure in the outer halo is
enhanced when these smoother components are excluded from the
clustering analysis (through a combination of the SDSS footprint and
the cut on $|Z|$). This is particularly true in the case of Aq-F,
where the majority of the mass in the halo is contributed by one
extensive and relatively `smooth' component. By contrast, in Aq-C the
average clustering amplitude \textit{decreases} on large scales when
we fix the Galactic plane. As noted above, in this case the massive
coherent streams that dominate the clustering signal of this halo
mostly fall outside the SDSS footprint.

Panel (c) of \fig{fig:sdss_mock_no} shows the randomly aligned
observations of panel (a) convolved with observational errors in
distance and velocity. These errors `smooth out' the halo, suppress
the clustering signal overall and increase the variance between
observers on small scales. Again the effect is most pronounced for
Aq-F, where blurring of the dominant smooth component further
decreases the contrast of substructure. In most cases these two
effects (alignment and observational errors) counteract each other to
produce the distribution of signals shown in
\fig{fig:sdss_mock_pa}. In the case of Aq-E the signal suffers
disproportionately from errors in the aligned configuration, perhaps
because this signal is due to a small number of pairs at large
distances.

\begin{figure*}
\includegraphics[height=65mm,clip=True]{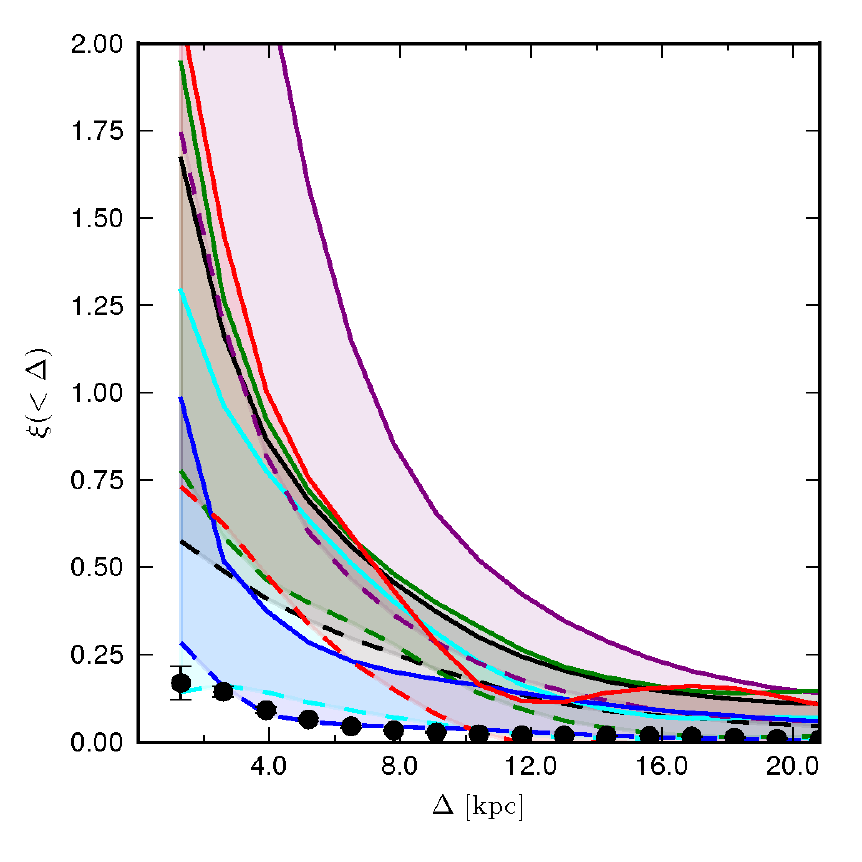}
\includegraphics[height=65mm,clip=True,trim=1.3cm 0 0 0]{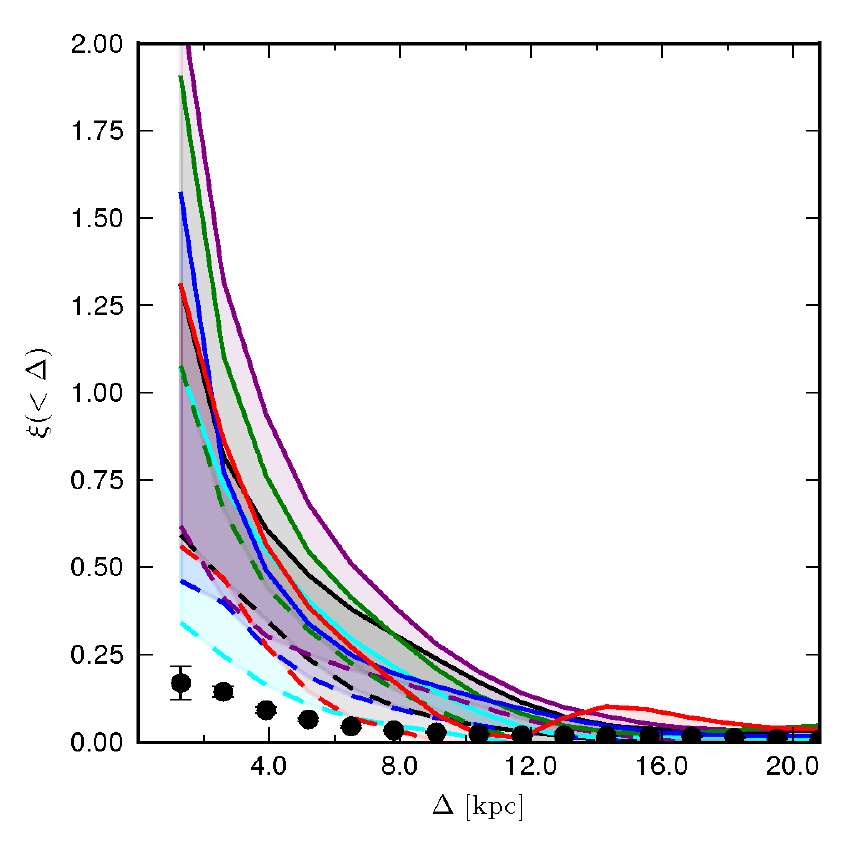}
\includegraphics[height=65mm,clip=True,trim=1.3cm 0 0 0]{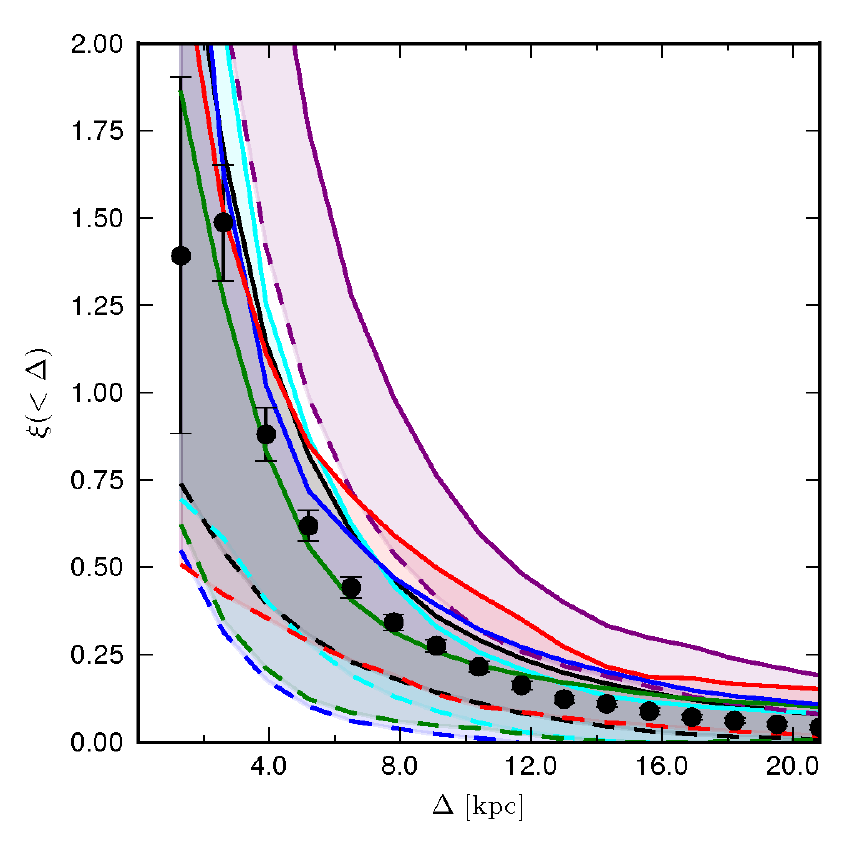}
\caption{\deltacf{} for mock observations, as
  \fig{fig:sdss_mock_pa}. Here we select three different
  galactocentric distance ranges in the halo: from $5<r<60$~kpc (left
  panel), from $5<r<20$~kpc (central panel) and from $30<r<60$~kpc
  (right panel). The Milky Way BHB clustering signal (black points) is
  recalculated for the distance range in each panel. Colours are as
  \fig{fig:sdss_mock_pa}.}
\label{fig:sdss_mock_distance}          
\end{figure*}

Finally, in \fig{fig:sdss_mock_distance} we examine differences
between nearby and more distant halo stars (also discussed by
Xue {et~al.} 2011). The left-hand panel corresponds to the full
range of the Xue {et~al.} (2008) sample ($5<r<60$~kpc), the central panel
corresponds to nearby stars ($5<r<20$~kpc) which we excluded in
\fig{fig:sdss_mock_pa}, and the right-hand panel corresponds to the
most distant stars in the sample ($30<r<60$~kpc). As shown in
\fig{fig:sdss_real}, including nearby stars considerably reduces the
amplitude of \deltacf{} for Milky Way halo BHBs, although the signal
in the lowest bins remains significantly above zero. The observational
data are dominated by stars at $r<20$~kpc and hence the Milky Way
signal is essentially identical in the left and central panels.

In the whole-halo sample (left-hand panel of
\fig{fig:sdss_mock_distance}) some of the simulations (particularly
Aq-A and Aq-B) show a reduction in \deltacf{} on small scales,
qualitatively similar to the observations. Nevertheless
\fig{fig:sdss_mock_distance} suggests that the contrast of structure
against smoothly distributed stars is too high in our models, and that
the discrepancy is most pronounced in the inner halo. Selecting an
inner-halo sample (central panel of \fig{fig:sdss_mock_distance})
emphasises this point. The inner halo shows a similar degree of
clustering in all six haloes, in all cases somewhat lower than the
outer halo signal (on intermediate scales), but much stronger than the
Milky Way data. This panel reinforces our comments above on the
structure of the haloes: Aq-C resembles Aq-A and Aq-D in the inner
halo, because this distance range excludes the massive coherent
streams that are the likely source of the strong signal seen in
\fig{fig:sdss_mock_pa}. Most of the coherent structure in Aq-E is
nearby and the inclusion of stars at greater distances
\textit{diminishes} the average signal. This is in contradiction to
the phase-mixing explanation for a more structured outer halo, but
Aq-E is unusual in that for $r<20$~kpc it is dominated by a highly
oblate, coherently rotating component (somewhat resembling a very
thick disc, but with a significant radial velocity dispersion). Aq-B
has a similar spherically averaged density profile to Aq-E, but here
the inner halo shows less correlation, as expected. In Aq-F \deltacf{}
is dominated by nearby stars, although these are highly clustered in
phase space.

The comparison with more distant stars shown in the right-hand
  panel of \fig{fig:sdss_mock_distance} suggest that most of the
  models are consistent with the data in this region of the halo. Only
  Aq-C is clearly discrepant (Aq-D is marginally inconsistent at
  scales below $\Delta\sim12$~kpc). That our simulations more closely
  match the outer regions of the halo (beyond $30$~kpc) supports the
  view that these regions are dominated by stars accreted from
  satellites (Xue {et~al.} 2011; Font {et~al.} 2011). The outer stellar halo
  of the Milky Way appears to be typical (or perhaps slightly more
  structured than average) with respect to our sample of six plausible
  realizations of dark matter haloes with similar mass.

\section{Conclusions}
\label{sec:conclusion}

We have analysed a correlation function for halo stars, \deltacf{},
defining their separation in four dimensions of phase space using a
metric (which we call $\Delta$) in readily-obtained observables
(angular and radial separation and radial velocity difference). A
statistic of this type usefully quantifies kinematic and spatial
substructure in the halo, and can be applied to observational data and
catalogues generated from theoretical models. This analysis is
particularly well suited to the distant halo -- other methods for
studying clustering in many dimensions may be more suitable for the
`fine grained' data on the nearby halo that will be obtained by the
\textit{Gaia} mission (e.g. G{\'{o}}mez {et~al.} 2010).

We have measured \deltacf{} for a large catalogue of spectroscopically
confirmed BHB stars from SDSS (Xue {et~al.} 2008). We find significant
clustering in these data, particularly when we restrict the sample to
stars with galactocentric distance $r>20$~kpc. This finding of
stronger phase space correlations between stars in the outer halo is
in agreement with that of Xue {et~al.} (2011). 

To test models of the accreted components of stellar haloes and
understand the effects of sample variance, we have computed \deltacf{}
for mock observations constructed from the six $\Lambda$CDM
simulations of Cooper {et~al.} (2010) in which only the stellar haloes
produced by disrupted satellites are considered. Our statistic
distinguishes quantitatively between these six qualitatively different
halo realizations. When only stars with $r>20$~kpc are considered,
five of our six simulations show statistically significant
correlations on scales in our metric equivalent to $\sim 1-8$~kpc (for
all observers on the Solar circle). Most of the models are consistent
with the Milky Way data for the outer halo, $r>30$~kpc. For the inner
halo, however, particularly at galactocentric distances smaller than
20~kpc, the simulations tend to be significantly more strongly
clustered than the data. One possible explanation for this is a
deficiency of smoothly distributed halo stars in the models, perhaps
attributable to the absence of so-called `in situ' halo stars. These
stars may be scattered from the Galactic disc, or born on eccentric
orbits (in streams of accreted gas or an unstable cooling flow, for
example). Neither of these processes are included in our model of the
accreted halo.

Although it seems reasonable to expect that in situ haloes are
distributed with spherical or axial symmetry and have a low degree of
coherence in phase space, models of such components and predictions
for the fraction of stars they contain are not well constrained. Most
hypotheses for in situ halo formation limit these stars to an `inner'
halo and predict that the accreted component (which we simulate)
dominates at larger radii
(e.g. Abadi {et~al.} 2006; Zolotov {et~al.} 2009; Font {et~al.} 2011). However, the fraction
of the halo formed in situ and the boundary between `inner' and
`outer' halo are highly model-dependent. Detections of observable
`dichotomies' in the Milky Way halo (Carollo {et~al.} 2007) are still
debated (e.g. Sch{\"{o}}nrich, Asplund \&  Casagrande 2011; Beers {et~al.} 2011). It is
possible to place broad limits on the fraction of stars in a `missing'
smooth component, for example by comparing the RMS variation of
projected star-counts in our models (Helmi {et~al.} 2011b) to the Milky
Way (Bell {et~al.} 2008). However, the uncertainties involved are substantial.

Another factor in the discrepancy between the models and the data may
be the absence of a baryonic (disc) contribution to the gravitational
potential. A massive disc could alter the process of satellite
disruption in the inner halo and might make the potential within
$30$~kpc more spherical (Kazantzidis, Abadi \&  Navarro 2010), possibly distributing
more inner halo stars into the SDSS footprint (on the other hand, a
more axisymmetric or spherical dark halo might also result in fewer
chaotic orbits, hence more coherent streams). Because of these
modelling uncertainties, our application of the \deltacf{} statistic
can presently serve only as a basic test for the abundance of
substructure in the simulations.

Several aspects of our approach could be improved. 
It seems desirable to use well-measured radial velocity data to
enhance clustering signals such as our correlation function relative
to those based on configuration space data alone. However, so far, no
proposal for 
including these velocity data is well-supported by theory. 
Here, we have used a straightforward choice of
parametrised metric to illustrate the concept of scaling radial
velocity separations to `equivalent' configuration space separations,
and this is empirically useful in recovering a measurable
signal. Nevertheless, we have not found any compelling or generic way
to select the scaling parameter
($w_{v}$). Improving either the definition of the metric itself or the
means of fixing this parameter is a clear priority for extensions of
this approach. A similar issue affects the weighting of velocity
information in clustering algorithms (e.g. Sharma {et~al.} 2010).

Finally, further comparisons between stellar halo models and
observational data should also account for the selection effects such
spectroscopic incompleteness and the potential bias of BHB stars as a
tracer of the stellar halo (Bell {et~al.} 2010; Xue {et~al.} 2011). For
statistical analysis, there is a pressing requirement for
observational samples with well-understood selection functions, even
if they do not probe the most distant halo. The LAMOST Galactic survey
is likely to be the first to approach this goal.

In summary, we have taken a first step in adapting a well-studied
cosmological statistic, the two-point correlation function, to the
study of the Milky Way halo. Our comparisons highlight the complexity
of statistical analysis in the stellar halo, and the importance of
interpreting observational results in the context of realistic models
of halo assembly. We have compared the SDSS data with the stellar
halos produced by disrupted satellites in {\em ab initio} galaxy
formation models constructed from the Aquarius N-body simulations of
galactic dark halos in the $\Lambda$CDM cosmology. With further
refinements and more data, our statistical approach to quantifying the
smoothness of the halo can provide a practical and productive way to
study the structure of the Milky Way halo and compare with theoretical
expectations.

\section*{Acknowledgements}

The authors thank Heather Morrison and the Spaghetti Survey team for
making their data available prior to publication, and Xiangxiang Xue
and Sergey Koposov for their assistance. We thank the referee for
their helpful comments which greatly improved the structure of this
paper. APC acknowledges an STFC studentship and thanks Else
Starkenburg for useful discussions. SMC acknowledges the support of a
Leverhulme Research Fellowship. CSF acknowledges a Royal Society
Wolfson Research Merit Award and ERC Advanced Investigator grant
267291- COSMIWAY. AH acknowledges funding support from the European
Research Council under ERC-StG grant GALACTICA-24027. This work was
supported in part by an STFC rolling grant to the ICC. The
calculations for this paper were performed on the ICC Cosmology
Machine, which is part of the DiRAC Facility jointly funded by STFC,
the Large Facilities Capital Fund of BIS, and Durham University.Fig. 1
was produced with the {\tt HEALPy} implementation of {\tt HEALPix}
[http://healpix.jpl.nasa.gov, G{\'{o}}rski {et~al.} 2005].

%\bibliographystyle{mn2e}
%% \bibliography
{}

\bsp

\label{lastpage}
\end{document}